# Bridging Finite Element and Molecular Dynamics for Non-Fourier Thermal Transport Near Nanoscale Hot Spot

Tanvirul Abedien, and Tianli Feng, *Member, IEEE*

*Abstract*—Nanoscale hot spots forming tens of nanometers beneath the gate in advanced FinFET and HEMT devices drive heat transport into a non-Fourier regime, challenging conventional (Fourier-based) finite-element (FEM) analyses and complicating the future thermal-aware chip design. Molecular dynamics (MD) naturally captures ballistic transport and phonon nonequilibrium, but has not been applied to hot-spot problems due to computational cost. Here, we try the first MD simulations of hot-spot heat transfer across ballistic–diffusive regimes and benchmark them against FEM. We find that FEM using bulk thermal conductivity $\kappa_0$ significantly underestimates hot spot temperature, even when the channel thickness is ~10 times of phonon mean free paths, indicating persistent non-Fourier effects. We introduce a size-dependent "best" conductivity, $\kappa_{best}$, using which FEM can reproduce MD hot-spot temperatures with high fidelity. We further decompose the MD-extracted thermal resistance into: (i) diffusive spreading, (ii) cross-plane ballistic, (iii) heat-carrier selective heating, and (iv) residual 3D ballistic-spreading resistances, and quantify each contribution. The resulting framework offers a practical route to embed non-Fourier physics into FEM for hot-spot prediction, reliability assessment, and thermally aware design of next-generation transistors.

This work was supported by the Nano & Material Technology Development Program through the National Research Foundation of Korea (NRF), funded by the Ministry of Science and ICT (RS-2024-00444574). (Corresponding author: Tianli Feng.)

Tanvirul Abedien and Tianli Feng are with the Department of Mechanical Engineering, University of Utah, Salt Lake City, Utah 84112, USA (e-mail: tanvirul.abedien@utah.edu; tianli.feng@utah.edu).

## I. INTRODUCTION

Miniaturization in semiconductor manufacturing has steadily increased the performance, energy efficiency, and integration density of integrated circuits (ICs). Modern processors now integrate more than 100 billion transistors; however, continued dimensional scaling has slowed, and thermal management has emerged as a central challenge for advanced electronics [1]. As future architectures place trillions of devices in tight proximity, thermal-aware design and innovative heat-removal strategies will be increasingly important. In field-effect transistors, dissipation occurs tens of nanometers beneath the gate, creating nanoscale "hot spots" with elevated local temperatures [2], [3], [4]. These hot spots degrade performance, perturb electrical characteristics, and undermine stable operation and long-term reliability [5], [6], [7]. Effective heat removal from these hot spots is therefore essential for ensuring the reliability and efficiency of nanoscale electronic systems [8].

Thermal transport near nanoscale hot spots cannot be accurately predicted using the traditional Fourier's law, which captures only diffusive heat flow and neglects ballistic effects and carrier nonequilibrium [4], [9], [10], [11]. When the phonon mean free path (MFP) becomes comparable to or exceeds the characteristic size of the materials, thermal transport becomes non-Fourier, leading to an increase in the hot spot temperature [12], [13]. The non-Fourier effects include (i) additional thermal resistance arising from ballistic phonons; (ii) a temperature discontinuity at the hot spot boundary; (iii) heat-carrier nonequilibrium both inside the hotspot and along the downstream path to the substrate due to the diverse MFPs of different phonons [14]. Moreover, Joule heating prefers to excite optical phonons than acoustic phonons, adding more to phonon non-equilibrium [7], [15], [16], [17], [18], [19], [20], which further contributes to an increased hot spot temperature since optical phonons have much lower capability to dissipate heat than acoustic phonons [19], [21]. Figure 1 illustrates the non-Fourier effects discussed above, which increase the hot spot temperature or equivalently decrease the effective thermal conductivity [22].

Significant efforts have been made in recent years to simulate non-Fourier thermal transport near nanoscale hot spots. The phonon Boltzmann transport equation (BTE) is the most widely used numerical method [23]. Originally, simplified BTE models—such as the gray approximation or asymptotic expansion of the spatial component of phonon distribution were employed [24], [25], [26], [27], [28] to investigate quasiballistic thermal transport near nanoscale hot spots, due to the limited computing power [29], [30]. These studies typically model nanoscale hot spots as equilibrium heat sources, with ballistic transport being the primary reason for variations in thermal conductivity, while phonon non-equilibrium effects near the hot spots are neglected [31], [32], [33]. Separately, Vallabhaneni et al. [19] proposed a multi-temperature model based on diffusive transport to study phonon non-equilibrium in a laser-heated graphene, later experimentally confirmed by Zobeiri et al. [34] and Dolleman et al [35]. However, ballistic phonon transport is neglected in these studies. To capture both ballistic transport and phonon non-equilibrium, Xu et al. implemented nongray phonon BTE to study the respective contributions of these effects in laser-heated hot spots in Raman experiments and Joule-heating in FinFET near nano hot spots [22]. Later, Sheng et al. incorporated non-Fourier effects in thermal generation into the BTE framework for a 3-D silicon-based FinFET, addressing the limitations of simplified assumptions or reliance on fitting parameters in thermal generation models, and achieved results that closely aligned with experimental observations [36]. Very recently, Rao et al. analyzed the influence of substrate thickness, heat source size, and substrate material on ballistic thermal resistance effects in FinFETs with nanoscale hot spots using the nongray BTE model [37]. In addition to BTE, Monte Carlo (MC) is an effective method to study non-Fourier transport [38]. For example, Nghiem et al. quantified the ballistic effect on phonon transport in silicon nano hot spots using Monte Carlo simulation [39], while Chen et al. combined a phonon BTE solver with electronic MC (e-MC) simulations to study thermal transport in a 1D silicon transistor and Si-based heterojunction interfaces, focusing on the coupling between hot spots and interfacial effects [21]. However, all these hot spot studies rely on the numerical solution of the phonon BTE, which is computationally expensive [40].

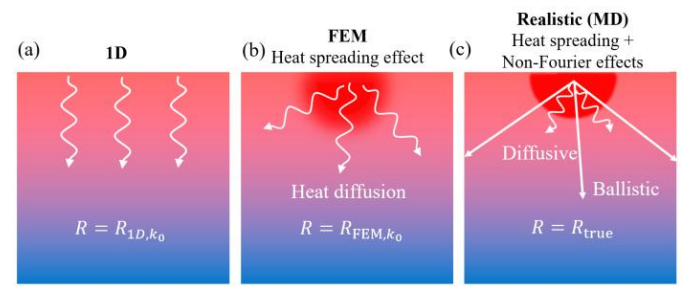

Fig. 1. Schematic temperature profiles in the fin of a FinFET or a channel layer of a HEMT for (a) 1D Fourier heat conduction, (b) 3D Fourier diffusion from a small hot spot into a wide substrate with a spreading resistance (3D effect), and (c) realistic 3D non-Fourier transport.

Molecular Dynamics (MD) is an alternative method to study non-Fourier thermal transport [41]. Compared to BTE, MD is more effective for simulating complex structures such as heterostructures and amorphous materials, and is capable of capturing all orders of anharmonicity and phonon nonequilibrium [40], [42]. However, MD simulations are constrained to domain sizes in the order of tens of nanometers for 3D geometries due to computational cost [41]. This size is much smaller than the dimensions of practical devices, as well as the phonon MFP in semiconductors such as Si, SiC, GaN, and diamond, which is above 100 nm, making MD infeasible to study the ballistic to diffusive transition. Thus, MD has not been used to study 3D thermal transport near nanoscale hot spots.

In this paper, we explore the possibility of using MD to simulate the nano hot spot thermal transport from ballistic to diffusive regimes. The results are compared to finite element method (FEM) simulations, which assume Fourier's thermal diffusion. With this comparison, we propose a new "best thermal conductivity", which can be used in FEM to reproduce the same temperature rise as in MD. In the end, we decompose the thermal resistance in MD into multiple components to understand the underestimation of Fourier's law.

## II. Methods

Normal semiconductors like Si, AlN, GaN, SiC, and diamond exhibit phonon MFPs in the order of hundreds of nanometers to several microns. Simulating their transition from ballistic to diffusive thermal transport in 3D domains thus demands simulation volumes exceeding 1 μm³, corresponding to more than 50 billion atoms, which lies far beyond practical MD limits. To explore both regimes within feasible computational costs, we instead adopt $SiO_2$ as a model system, which exhibits an average phonon MFP ($\Lambda_{ph}$) of approximately 12 nm based on our non-equilibrium MD simulations; taking the diffusive limit as > 10×MFP yields a domain size of ~120 nm, which is readily simulated via MD. Note that $SiO_2$ is used here purely as a representative model system to quantify the relative error between FEM and MD predictions.

The conclusions drawn in this paper can be extended to other materials such as Si, AlN, GaN, SiC, and diamond, since all the properties are nondimensionalized to a relative quantity. For example, the device length is relative to the phonon mean free path, thermal resistance is relative to the FEM, and thermal conductivity is relative to the bulk value. To extend these conclusions to other materials (e.g., Si, AlN, GaN, SiC), simply scale the device dimensions by the ratio of their phonon mean free path and scale the thermal conductivities by the bulk value. The error of this scaling from one material to the other material should be small as long as the effective phonon mean free path can well capture the size-dependent thermal conductivity.

Both FEM and MD use an identical simulation domain to model transistor structures representative of FinFET/HEMT devices, as shown in Fig. 2. To span ballistic to diffusive regimes, the channel layer thickness $t$ is varied from $0.25\Lambda_{ph}$ to $12.5\Lambda_{ph}$, corresponding to Knudsen number ($Kn_t = \Lambda_{ph}/t$) from 4 (ballistic) to 0.08 (diffusive). The width is $W = 1.25\Lambda_{ph}$, with periodic lateral boundaries. Due to computational cost, we are not able to simulate a domain with both large thickness $t$ and width $W$. The hot spot is located at the top center with a width $W_{hot}$ varying from $0.13\ W$ to $1\ W$ and a depth of $0.2\Lambda_{ph}$. The length is $L = 0.36\ \Lambda_{ph}$, with an adiabatic boundary condition applied. The bottom of the domain is maintained at a constant temperature of 300 K.

All MD simulations begin with 0.5 ns equilibration under constant volume and temperature (NVT) at 300 K using a 0.5 fs time step. The stabilized systems are then switched to a constant energy and volume (NVE) ensemble with a constant heat input of 2.0714 eV/ps imposed in the hot spot, while the bottom boundary is held at 300 K. After a 2 ns transient to establish a steady temperature gradient, the temperature and heat-flux data are collected over an additional 2 ns for analysis.

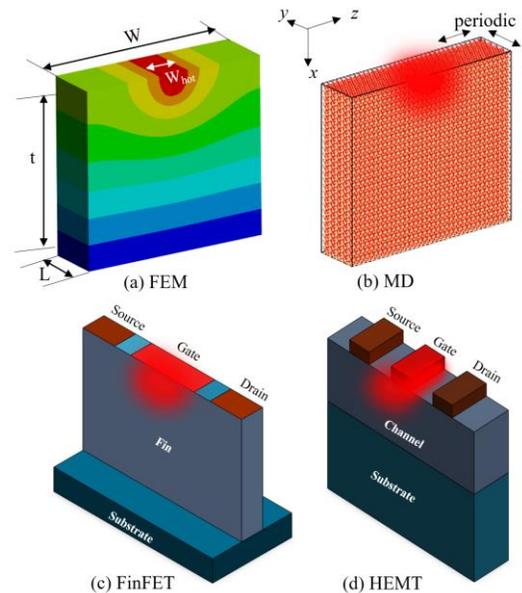

Fig. 2. (a) FEM and (b) NEMD domain configurations used for both (c) FinFET and (d) HEMT structures to simulate nanoscale hot spots.

All FEM simulations are performed in ANSYS 2024 R1. The computational domain is discretized into a $10^5$ mesh after a convergence test. For all simulations, thermal conductivity is assumed constant, independent of temperature, and internal heat generation is prescribed as a boundary condition to mimic the hot spot. For both MD and FEM, the total thermal resistances are defined as:

$$R_{\text{total}} = \frac{\Delta T \cdot A_c}{\dot{Q}} \quad (1)$$

where $A_c$ is the horizontal cross-sectional area of the substrate, $\Delta T = T_{\text{hot spot}} - T_{\text{bottom}}$, and $\dot{Q}$ is the heating power.

## III. RESULTS

### A. Total resistances given by MD vs FEM

In Fig. 3, we plot the thermal resistance obtained from both MD and FEM for channel layer thickness $t$ from $0.25\Lambda_{\text{ph}}$ up to $12.5\Lambda_{\text{ph}}$, covering the full range from ballistic to diffusive transport. For each $t$, the hot spot width $W_{\text{hot}}$ is varied systematically from $0.13\,W$ up to the full device width $W$. FEM consistently underpredicts the thermal resistance given by MD, with the largest discrepancies occurring when the device length is smallest. For example, at $t = 0.25\,\Lambda_{\text{ph}}$, FEM yields only about 30% of the resistance computed by MD, meaning the actual resistance is roughly ~3.5 times higher, while at $t \approx 1\,\Lambda_{\text{ph}}$, the real resistance remains ~2.5 times larger than the FEM prediction. Even when $t$ reaches $12.5\,\Lambda_{\text{ph}}$, FEM still underestimates resistance by ~15%, a nonnegligible error that gradually vanishes only once $t$ extends well beyond the phonon MFP $\Lambda_{\text{ph}}$.

The discrepancy between FEM and MD arises from non-Fourier effects, which can be directly measured by the ratio $R_{\text{MD}}/R_{\text{FEM},\kappa_0}$. As shown in Fig. 4(a), the results are plotted at various Knudsen numbers of thickness ($Kn_t = \Lambda_{\text{ph}}/t$) and hot spot width ($Kn_w = \Lambda_{\text{ph}}/W_{\text{hot}}$). In the ballistic regime ($t = 0.25\,\Lambda_{\text{ph}}$), non-Fourier effects amplify the thermal resistance by roughly 3.5 times, independent of hot spot localization ($W_{\text{hot}}/W$). Even deep in the diffusive regime ($t = 12.5\,\Lambda_{\text{ph}}$), the non-Fourier effects still add about 10–20 % extra resistance, demonstrating their continued relevance. When instead we compare the MD resistance with the one-dimensional Fourier baseline, $R_{1D,\kappa_0} \equiv t/\kappa_0$, the ratio $R_{\text{MD}}/R_{1D,\kappa_0}$ captures both non-Fourier transport and lateral heat-spreading resistance (Fig. 4 b). Because heat spreading is more pronounced for narrower hot spots, the combined effect reaches roughly 7.5× the 1D-Fourier value in our thinnest device ($t = 0.25\,\Lambda_{\text{ph}}$) with the tightest hot spot ($W_{\text{hot}}/W = 0.125$).

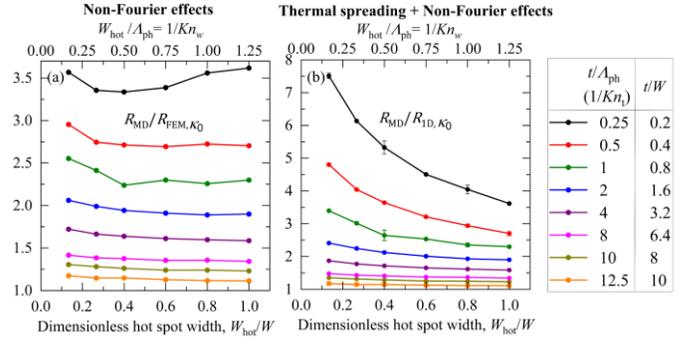

Fig. 4. Thermal resistance ratios $R_{MD}/R_{FEM,\kappa_0}$ (a) and $R_{MD}/R_{1D,\kappa_0}$ (b) as a function of dimensionless hot spot width ($W_{hot}/W$ or $W_{hot}/\Lambda_{ph}$) at various channel layer thicknesses $t$.

### B. Best thermal conductivities for use in FEM

Because FEM remains the industry standard for heat-transfer modeling, offering ease of use, established workflows, and graphical interfaces, it is preferable to retain FEM rather than adopt the computationally intensive and less accessible MD approach. To reconcile the FEM predictions with MD benchmarks for both overall resistance and hot spot temperature, we therefore introduce an effective thermal conductivity, $\kappa_{\text{best}}$, which can be used in FEM to replace the bulk value $\kappa_0$. With $\kappa_{\text{best}}$, FEM can reproduce the same temperature rise as in MD. In other words, one can continue using FEM's familiar framework (with $\kappa_{\text{best}}$) to capture the non-Fourier and size-dependent effects. Figure 5 plots the fitted best values of dimensionless $\kappa_{\text{best}}/\kappa_0$ against normalized device thickness ($t/\Lambda_{\text{ph}}$) and hot spot width ($W_{\text{hot}}/W$). In every case $\kappa_{\text{best}} < \kappa_0$, and it exhibits a pronounced size dependence: for $t = 0.25\, \Lambda_{\text{ph}}$, $\kappa_{\text{best}} \approx 0.3\, \kappa_0$, whereas even at $t = 12.5\, \Lambda_{\text{ph}}$, it remains suppressed to about 0.85–0.9 $\kappa_0$ (roughly 10% below the bulk value).

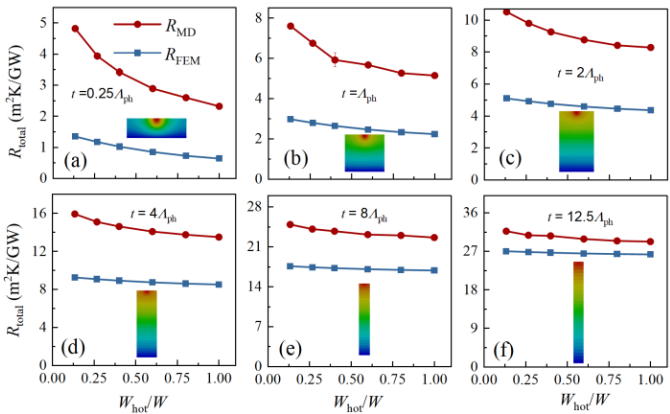

Fig. 3. Total thermal resistances given by FEM vs MD at various normalized hot spot sizes "$W_{hot}/W$" and transistor channel layer thicknesses "$t/\Lambda_{ph}$".

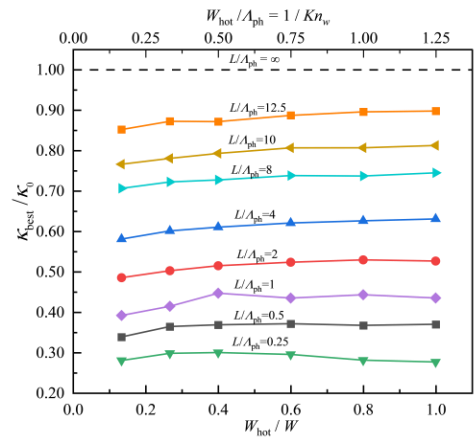

Fig. 5. Fitted "best thermal conductivity" $\kappa_{best}$ that can be used in FEM to reproduce the MD hot spot temperature.

Because $\kappa_{\text{best}}$ is defined nondimensionally, Fig. 5 can be transplanted to semiconductor materials such as Si, GaN, AlN, SiC, and diamond with small additional errors. For instance, silicon's phonon MFP spectrum spans 10 nm to 50 μm, with a cumulative-conductivity midpoint near 400 nm; taking 400 nm as the average MFP [43], Fig. 5 indicates that FEM should use an effective conductivity of roughly 0.5 $\kappa_{\text{Si},0}$ for a 1 μm-thick device, 0.65 $\kappa_{\text{Si},0}$ for 2 μm, and 0.9 $\kappa_{\text{Si},0}$ for 5 μm (with even lower values if the hot spot is much narrower than the MFP). GaN's MFP midpoint is about 300 nm [44], corresponding to $\kappa_{\text{best}} \approx 0.55\, \kappa_{\text{GaN},0}$, $0.7\, \kappa_{\text{GaN},0}$, and $0.95\, \kappa_{\text{GaN},0}$ for channel layer thicknesses of 1, 2, and 5 μm, respectively; AlN and SiC follow similar trends. Diamond, with MFPs from ~400 nm to 40 μm and a midpoint near 700 nm [43], would require $\kappa_{\text{best}} \approx 0.45\, \kappa_{\text{diamond},0}$, $0.55\, \kappa_{\text{diamond},0}$, and $0.7\, \kappa_{\text{diamond},0}$ for 1, 2, and 5 μm devices. Note that these values are rough estimation considering that MFP is not a single value but has a spectrum, which will affect $\kappa_{\text{best}}/\kappa_0$ from materials to materials.

Figure 6 compares the MD temperature profile with FEM predictions using either the bulk conductivity $\kappa_0$ or the calibrated $\kappa_{\text{best}}$. With $\kappa_0$, FEM substantially underpredicts the hot spot temperature, whereas substituting $\kappa_{\text{best}}$ brings the FEM hot spot temperature into excellent agreement with MD. This agreement is localized, however: even with $\kappa_{\text{best}}$, FEM cannot reproduce the characteristic temperature jump at the heat-source boundary seen in MD, reflecting FEM's inherent inability to model non-Fourier transport. Nevertheless, this limitation does not impede accurate hot spot prediction. Because the device's peak temperature (and thus its reliability and failure risk) is determined by the hot spot, FEM with $\kappa_{\text{best}}$ is sufficient for lifetime and risk assessments.

## IV. Discussion: Total Resistance Decomposition

To quantify the non-Fourier effects in MD and understand the error of FEM using $\kappa_0$, we decompose the total resistance from the hot spot to the substrate. FEM resistance can be decomposed into

$$R_{\text{FEM},\kappa_0} = R_{\text{1D},\kappa_0} \cdot r_{\text{sprd,diff}} \qquad (2)$$

where $R_{\text{1D},\kappa_0} = t/\kappa_0$ is the thermal resistance assuming heat transfer from hot spot to the substrate is 1D, and $r_{\text{sprd,diff}}$ is the spreading resistance from the small area of hot spot to the large area of the substrate in the diffusive limit. In contrast, the true thermal resistance considering non-Fourier effects has been proposed to be in the literature

$$R_{\text{true}}^{\text{literature}} = R_{\text{FEM},\kappa_0} \cdot r_{\text{1D,b}} \cdot r_{\text{spot,b}} \qquad (3)$$

by using BTE and MC simulations [45], [46], [47]. Here, the dimensionless factor $r_{\text{1D,b}}$ measures the cross-plane ballistic effect due to the thin thickness $t$ compared to the phonon MFP, and $r_{\text{spot,b}}$ is the ballistic effect due to the small dimension of the hot spot compared to the phonon MFP. These factors could explain well the temperature obtained from BTE and MC simulations. In addition to them, in real transistors, hot electrons heat optical phonons first, which then transfer energy to acoustic phonons. Therefore, the phonon heat carriers are selectively heated, rather than uniformly heated. Thus, we add one more term $r_{\text{sel,heat}}$ to account for this selective heating effect, and the total resistance becomes

$$R_{\text{true}} = R_{\text{MD}} = R_{\text{FEM},\kappa_0} \cdot r_{\text{1D,b}} \cdot r_{\text{sel,heat}} \cdot r_{\text{res,b}} \qquad (4)$$

We also replace $r_{\text{spot,b}}$ with a residual ballistic effect, $r_{\text{res,b}}$, as explained in the Sec. IV.D. The amplitude of each term on the right-hand side of Eq. (4) is discussed in the following sections.

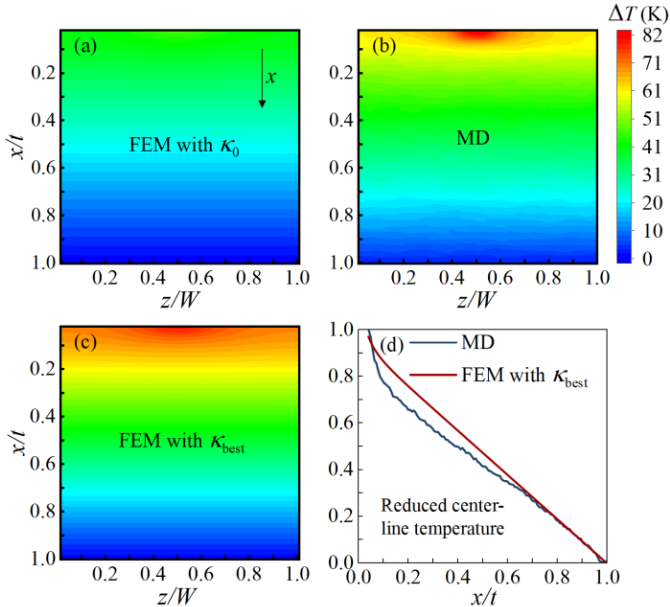

Fig. 6. Temperature profile of the domain ($L/W$=3.2, $L/\Lambda_{ph}$ =4) when simulated with (a) FEM using $\kappa_0$, (b) MD, and (c) FEM using $\kappa_{\text{best}}$. (d) Non-dimensional temperature rise along the center line.

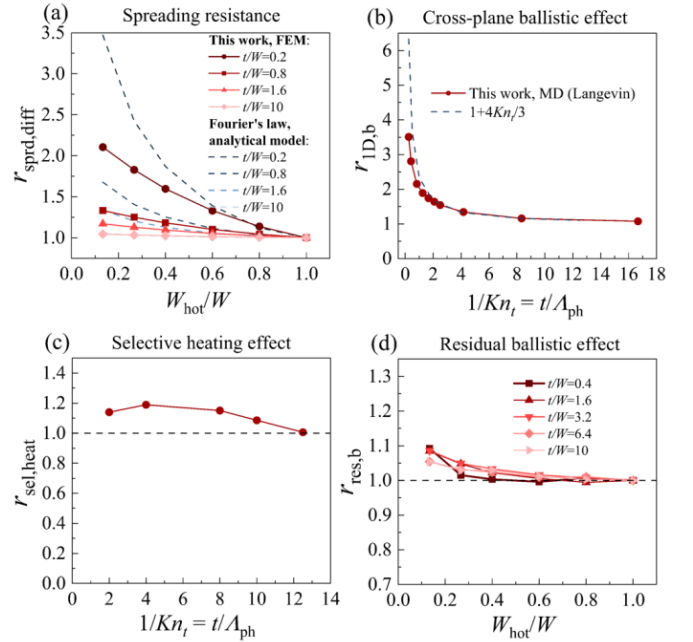

Fig. 7. (a) Diffusive thermal spreading resistance simulated by FEM. (b) Cross-plane ballistic effect simulated by NEMD. (c) Selective heating-induced resistance simulated by NEMD. (d) Residual resistance due to the spreading of ballistic phonons.

The total non-Fourier effect, $R_{\text{MD}}/R_{\text{FEM},\kappa_0} = r_{\text{1D,b}} \cdot r_{\text{sel,heat}} \cdot r_{\text{res,b}}$ has been plotted in Fig. 4 (b). This effect can reach 3.5 for thin devices (e.g., $t = 0.25\, \Lambda_{\text{ph}}$) and is about 1.2 even when device is 10 times of phonon MFP.

### A. Diffusive thermal spreading resistance

Diffusive thermal spreading resistance, $r_{\text{sprd,diff}}$, is defined as the ratio of the 3D resistance to the 1D Fourier conduction resistance, i.e., $R_{\text{FEM},\kappa_0}/R_{\text{1D},\kappa_0}$, which becomes important when heat is transferred from a small heat source to a larger area [48], [49], [50]. As shown in Fig. 7(a), $r_{\text{sprd,diff}}$ is significant when both the device thickness-to-width ratio $t/W$ and the normalized hot spot width $W_{\text{hot}}/W$ are small, i.e., for thin devices and small hot spots. If either parameter is large, the spreading resistance is negligible regardless of the other. For example, in a thin device (e.g., $t/W = 0.2$), $r_{\text{sprd,diff}}$ can reach ~2.0 for a small hot spot ($W_{\text{hot}}/W = 0.2$) and ~1.5 for a medium hot spot ($W_{\text{hot}}/W = 0.5$). In contrast, for a thick device (e.g., $t/W = 10$), the spreading resistance is negligible, i.e., $r_{\text{sprd,diff}} \approx 1$, for all hot spot sizes.

The FEM-derived $r_{\text{sprd,diff}}$ values are compared with an analytical expression based on Fourier's law [48]

$$r_{\text{sprd,diff}} = 1 + \left(\frac{W}{W_{\text{hot}}}\right)^2 \left(\frac{W}{t}\right) \sum_{n=1}^{\infty} \frac{8 \sin^2\left(\frac{W_{\text{hot}} n\pi}{2W}\right) \cdot \cos^2\left(\frac{n\pi}{2}\right)}{(n\pi)^3 \coth\left(\frac{Ln\pi}{W}\right)}, \quad (5)$$

showing good qualitative agreement. For thin devices (smaller $t/W$), however, the simulated $r_{\text{sprd,diff}}$ is lower than the analytical prediction. This difference stems from the heat source model: the analytical formula assumes a constant heat-flux boundary condition, whereas the FEM uses uniform volumetric heat generation. As shown by Li et al. [46], a constant heat-flux boundary condition is equivalent to injecting heat over an infinitely thin region, which yields to higher spreading resistance than a finite-volume volumetric source. The volumetric model more closely reflects realistic transistor heating. The discrepancy between the two models diminishes as $t/W$ increases.

### B. Cross-plane Ballistic effect

The cross-plane ballistic factor, $r_{\text{1D,b}}$, measures the additional resistance caused by the ballistic transport of phonons along the vertical (hot spot-to-substrate) direction [51]. By definition, $r_{\text{1D,b}} = \frac{R_{\text{1D},t}}{R_{\text{1D},\kappa_0}} = R_{\text{1D},t} \cdot \frac{\kappa_0}{t}$, where $R_{\text{1D},t}$ is one-dimensional resistance from the hot spot to the substrate when the hot spot area is expanded to the entire cross-section, enforcing purely cross-plane heat flow with no lateral component. $R_{\text{1D},t}$ is obtained by NEMD simulations using a Langevin thermostat [52], which is employed to avoid mode-selective heating: whereas other thermostats can preferentially excite certain phonon modes, the Langevin approach injects energy uniformly across modes. Figure 7(b) shows $r_{\text{1D,b}}$ as a function of $t/\Lambda_{\text{ph}}$, compared to the empirical gray model [11]:

$$r_{\text{1D,b}} = 1 + \frac{4}{3} Kn_t. \quad (6)$$

They show good matches with each other. The ballistic transport limits the MFP of phonons and creates additional thermal resistance. Even when the device thickness is 10 times of phonon MFP, $r_{\text{1D,b}}$ is still around 1.13, being non-negligible.

### C. Selective heating effect

In practical transistors, hot electrons first excite optical phonons, which subsequently transfer energy to acoustic phonons. Consequently, the phonon system is heated selectively rather than uniformly. Because optical phonons contribute far less than acoustic phonons to the lattice thermal conductivity, preferentially heating optical modes lowers the effective thermal conductivity, introducing an additional resistance quantified by $r_{\text{sel,heat}}$, defined as

$$r_{\text{sel,heat}} = R_{\text{1D},t,\text{sel heat}}/R_{\text{1D},t,\text{unif heat}} \quad (7)$$

where $R_{\text{1D},t,\text{sel heat}}$ and $R_{\text{1D},t,\text{unif heat}}$ are the resistances of a thin film with thickness $t$ when the heat sources excite phonon modes selectively and uniformly, respectively. Depending on how the phonons are selectively excited in the heat source, $R_{\text{1D},t,\text{sel heat}}$ can be various values. Here, we just pick one example in MD to illustrate the effect, and the real effect in real resistors depends on the real operating condition and materials in transistors. In MD, examples of selective and uniform heating reservoirs are Nose-Hoover and Langevin thermostats, respectively [53], [54]. The former excites more optical phonons than acoustic phonons. The obtained $r_{\text{sel,heat}} = R_{\text{1D},t,\text{Nose}}/R_{\text{1D},t,\text{Langevin}}$ is shown in Fig. 7(c). At a broad range of device thickness ($t \leq 10\, \Lambda_{\text{ph}}$), $r_{\text{sel,heat}}$ contribute 10-20% additional thermal resistance. Note that the obtained $R_{\text{1D},t,\text{Nose}}$ values match well with full-band MC simulations by Shen et al. [47], though minor discrepancies remain due to differences in material properties. Our analysis indicates that selective excitation has a nonnegligible contribution to total thermal resistance specially when device is in ballistic limit, a phenomenon that has not been demonstrated robustly in previous studies.

### D. Residual ballistic effect

After applying all previously discussed factors $r_{\text{sprd,diff}} \cdot r_{\text{1D,b}} \cdot r_{\text{sel,heat}}$ to $R_{\text{1D},\kappa_0}$, the result still deviates from $R_{\text{MD}}$. The reason is that the "3D ballistic phonon spreading resistance" in MD is not equivalent to a "3D diffusive spreading resistance" plus a "1D ballistic resistance". Thus, there is an additional "residual" resistance associated with ballistic phonons spreading from the hot spot into the substrate, which can be obtained by:

$$r_{\text{res,b}} = \frac{R_{\text{MD}}}{R_{\text{1D},\kappa_0}\, r_{\text{sprd,diff}}\, r_{\text{1D,b}}\, r_{\text{sel,heat}}} \quad (8)$$

Since the physical meaning of $r_{\text{res,b}}$ is related to the ballistic transport of phonon along all directions out from the hot spot, it should depend on the distances from the hot spot to edges and substate. We note that in the literature [55], a different factor $r_{\text{spot,b}}$ was used instead of $r_{\text{res,b}}$, and it was interpreted as the intra-hot spot phonon ballisticity, which depends on $Kn_a =$

$\Lambda_{\text{ph}}/a$ and $Kn_b = \Lambda_{\text{ph}}/b$ (where $a$ and $b$ are the width and length of the hot spot). This description does not account for the extra resistance induced by the ballistic spreading out of the hot spot, and therefore, we consider $r_{\text{res,b}}$ is a more appropriate description. As shown in Fig. 7(d), $r_{\text{res,b}}$ is small in most cases and becomes noticeable only for very thin devices with very small hot spots. At $t/W = 0.4$ and $W_{\text{hot}}/W = 0.1$, it contributes ~10% additional resistance.

## V. Conclusions

In summary, we have successfully employed NEMD to simulate nanoscale hot spot heat transfer from ballistic to diffusive limit. MD temperature rises are found to be much higher than those of FEM due to non-Fourier effects. Using MD, we calibrate an effective "best" conductivity $\kappa_{\text{best}}$; substituting $\kappa_{\text{best}}$ in FEM reproduces the MD hot spot temperature rise, bridging the FEM-MD gap and capturing non-Fourier effects within standard FEM workflows. We further decomposed the MD-extracted total thermal resistance into four contributions: (i) diffusive spreading from a small source to a larger substrate area, (ii) cross-plane ballisticity associated with finite thickness and phonon boundary scattering, (iii) selective heating of optical phonons that drives carrier nonequilibrium, and (iv) a residual 3D ballistic-spreading term. This decomposition explains why $\kappa_0$-based FEM fails and identifies where corrections are most consequential. Because the results are nondimensionalized, the guidance is portable to common semiconductor platforms. The proposed $\kappa_{\text{best}}$ maps offer a practical, drop-in route to embed non-Fourier physics in FEM for hot-spot prediction, reliability assessment, and thermally aware design of next-generation electronics.

## Acknowledgment

We thank the support from Center for High Performance Computing (CHPC) at the University of Utah.